\documentstyle[12pt]{article}    
\begin{document}     
\thispagestyle{empty}     
\begin{flushright}    
UA/NPPS-1-05\\    
\end{flushright}    
    
\begin{center}    
{\large{\bf    
THE CRITICAL ENDPOINT OF\\    
BOOTSTRAP AND LATTICE QCD MATTER\\}}    
\vspace{2cm}     
{\large N. G. Antoniou$^{1,2}$, F. K. Diakonos$^1$ and A. S. Kapoyannis$^1$*}\\     
\smallskip     
{\it $^1$Department of Physics, University of Athens, 15771 Athens, Greece\\     
     $^2$Nuclear Theory Group, Brookhaven National Laboratory, Upton,    
 NY 11973, USA}    
\vspace{1cm}

\end{center}    
\vspace{0.5cm}    
\begin{abstract}    
The critical sector of strong interactions at high temperatures is explored    
in the frame of two complementary approaches: Statistical Bootstrap for the    
hadronic phase and Lattice QCD for the Quark-Gluon partition function.    
A region of thermodynamic instability of hadronic matter was found, as a    
direct prediction of Statistical Bootstrap. As a result, critical endpoint    
solutions for nonzero chemical potential were traced in the phase diagram    
of strongly interacting matter.    
These solutions are compared with recent lattice QCD results and their    
proximity to the freeze-out points of experiments with nuclei at high    
energies is also discussed.    
\end{abstract}    
    
\vspace{3cm}    
    
    
*Corresponding author.    
{\it E-mail address:} akapog@cc.uoa.gr (A. S. Kapoyannis)

\newpage    
\setcounter{page}{1}    
    
{\large \bf 1. Introduction}    
    
Quantum Chromodynamics is unquestionably the fundamental theory of strong    
interaction. However, the nonperturbative aspects of QCD belong still to a    
field of intense investigation. In fact, most of the properties of the    
hadronic world cannot be extracted yet from first QCD principles and the    
partition function of interacting hadrons produced and thermalized in    
high-energy collisions can only be determined within specific models of    
varying degree of limitation.    
    
In this paper we employ Statistical Bootstrap, including volume corrections    
for the finite size of hadrons, in order to describe the hadronic phase. The    
virtue of this description is associated with the fact that, among Statistical    
models of hadrons, it is the only one which predicts the onset of a phase    
transition in strongly interacting matter and therefore it is compatible with    
the properties of the QCD vacuum at high temperatures [1].    
The aim of this work is to pursue the search for this compatibility with QCD    
even further, asking whether Statistical Bootstrap of hadrons is consistent    
with the existence of a critical endpoint in strongly interacting matter, at    
high temperature and finite (nonzero) baryonic chemical potential. The    
existence of such a critical point in the phase diagram is required by QCD    
in extreme conditions [2,3] and it is the remnant of chiral QCD phase transition    
[2,3].    
    
The basic ingredients in our approach are (a) the hadronic partition function    
extracted from the equations of Statistical Bootstrap and (b) the equation of    
state of the quark-gluon phase given by recent QCD studies on the lattice    
[4-7]. Our principal aim from the matching of these two descriptions is to trace    
the formation of a critical point in the quark-hadron phase transition with a    
mechanism compatible both with Statistical Bootstrap and Lattice QCD.    
   
In [8,9] a similar scheme was pursued but with the quark-gluon partition 
function meditated by the MIT bag model.  
This model is a crude approximation leading either to an ideal gas of quarks
and gluons [8] near the critical line or to a gas of weakly interacting quarks
and gluons with perturbative terms in the sector of two light flavours [9]. 
The only non perturbative effect in this model comes from the pressure of the 
vacuum through the phenomenological bag constant. In order to have an adequate 
description of QCD matter in this approach the use of lattice equation of state, 
derived from first principles is an essential improvement. To this end we have 
employed, in this work, a realistic partition function of QCD matter which has 
become available recently as a result of a breakthrough achieved in lattice QCD 
studies allowing for computations at nonzero chemical potential [4,7]. As far as 
the hadronic phase is concerned we continue to use in this work the pressure 
partition function but with the volume Laplace conjugate variable   
$\xi$ as an active thermodynamic variable and not fixed at the zero value.   
These calculations, within a three flavour bootstrap scheme, are used here for   
the first time. Despite all these alterations we shall show that the existence   
of a critical point at finite chemical potential persists and we shall try to   
narrow the area of its location.    
   
In Section 2 the principles and basic equations of Statistical Bootstrap,    
leading to the partition function of hadronic phase, are briefly summarized    
and the appearance, within this framework, of a thermodynamic instability is    
discussed in detail. This instability is the origin of the formation of a    
critical point in the bootstrap matter. In Section 3 the partition function    
of the quark-gluon phase is extracted from recent lattice calculations of the    
pressure of QCD matter [4]. In Section 4 the above two descriptions of    
strongly interacting matter are exploited in a search for a critical point    
in the phase diagram, the location of which is fixed by a set of equations    
(29-33). Finally, in Section 5 our conclusions    
are presented whereas in the Appendix certain technical    
points are discussed concerning the evaluation of the quark-gluon partition    
function (part A) and the complete form that acquires the equation of maximum    
hadronic pressure (part B).

\vspace{1cm}    
    
{\large \bf 2. Thermodynamic Instability in Statistical Bootstrap}    
    
The main attribute of the Statistical Bootstrap (SB) [10-13] is that it    
describes thermodynamically interacting hadronic systems.    
Generally as far as its thermodynamic behaviour is concerned a system of    
strongly    
interacting entities may be considered as a collection of particles with    
the explicit form of the complex interaction among themselves. Another way    
to deal with the problem is to assume that the strongly interacting entities    
form a greater entity with certain mass. Then, if this mass is known, the    
greater entities may be treated as non-interacting and the simple    
thermodynamic description of an ideal system of particles may be applied.    
The SB adopts the second approach and assumes that the strongly interacting    
hadrons form greater clusters, called ``fireballs''. The problem then is    
moved to determine the mass of these clusters or to evaluate their mass    
spectrum $\tilde{\tau}(m^2)dm^2$, which is equal to the number of    
discrete hadronic states in the mass interval $\{m,m+dm\}$. The solution is    
based on the assumption that the fireballs may consist of smaller fireballs    
with the {\it same} mass spectrum or ``input'' particles which may not be    
divided further. The integral bootstrap equation then reads [14-17]    
    
\[    
\tilde{B}(p^2)\tilde{\tau}(p^2,b,s,\ldots)=    
\underbrace{g_{b,s,\ldots}\tilde{B}(p^2)\delta_0(p^2-m_{b,s,\ldots}^2)}_    
{input\;term}    
+\sum_{n=2}^\infty\frac{1}{n!}\int\delta^4\left(p-\sum_{i=1}^np_i\right)    
\cdot    
\]    
\begin{equation}     
\cdot\sum_{\{b_i\}}\delta_K\left(b-\sum_{i=1}^n b_i\right)     
\sum_{\{s_i\}}\delta_K\left(s-\sum_{i=1}^n s_i\right) \ldots     
\prod_{i=1}^n\tilde{B}(p_i^2)\tilde{\tau}(p_i^2,b_i,s_i,\ldots)d^4p_i\;,     
\end{equation}     
where    
$\delta_0(p^2-m_{b,s,\ldots}^2)=\delta(p^2-m_{b,s,\ldots}^2)\theta(p_0)$.    
Equation (1) also imposes conservation of four-momentum $p$ and    
any kind of existing additive quantum number (baryon number $b$, strangeness    
$s$, etc.), through the Kronecker function $\delta_K$, between a fireball and    
its constituent fireballs.    
The masses $m_{b,s,\ldots}$ are the ``input'' particle masses which    
constitute the smaller fireballs that can be formed and $g_{b,s,\ldots}$ are    
degeneracy factors due to the spin of the ``input'' particles.    
The correct counting of states of a fireball involves,    
apart from its mass spectrum which is of dynamical origin,    
a kinematical term $\tilde{B}(p^2)$. This term is related to    
the appropriate volume of a fireball which is considered to be carried with it,    
$V^{\mu}=\frac{V}{m}p^{\mu}$.    
The imposition of the requirement that the sum of volumes of the constituent    
fireballs has to be equal to the volume of the large fireball, as well as,    
momentum conservation lead to the fact that all fireballs possess the same    
volume to mass ratio. This ratio can be connected to the MIT bag constant,    
$\frac{V}{m}=\frac{1}{4B}$. The term $\tilde{B}(p^2)$, appearing in eq.~(1)    
is then    
\begin{equation}    
\tilde{B}(p^2)=\frac{2V^{\mu}p_{\mu}}{(2\pi)^3}=\frac{2Vm}{(2\pi)^3}\;.    
\end{equation}    
    
Eq.~(2) corresponds to a specific choice for $\tilde{B}$.   
Since in (1) the mass spectrum $\tilde{\tau}$ is accompanied by the term    
$\tilde{B}$, the bootstrap equation (apart from the input term) remains    
unchanged if $\tilde{\tau}$ and $\tilde{B}$ are redefined in such way so    
\begin{equation}   
\tilde{B}\tilde{\tau}=\tilde{B}'\tilde{\tau}'   
\end{equation}    
Every such transformation leads to different    
thermodynamic description of the system of fireballs, since the relevant    
quantity for this description is only the mass spectrum $\tilde{\tau}'$ and    
not the term $\tilde{B}'$. These transformations are not uniquely    
determined in the general bootstrap scheme, allowing for various versions    
of the model [15].    
    
The bootstrap equation can be simplified after performing a series of    
Laplace transformations to acquire the form    
\begin{equation}     
\varphi(\beta,\{\lambda\})=2G(\beta,\{\lambda\})     
-\exp [G(\beta,\{\lambda\})]+1\;.     
\end{equation}    
In the last equation $G(\beta,\{\lambda\})$ is the Laplace transform of the    
mass spectrum with the accompanying kinematical term    
\[    
G(\beta,\{\lambda\})=    
\sum_{b'=-\infty}^{\infty}\lambda_b^{b'}    
\sum_{q'=-\infty}^{\infty}\lambda_q^{q'}    
\sum_{s'=-\infty}^{\infty}\lambda_S^{s'}    
\sum_{|s'|=0}^{\infty}\gamma_s^{|s'|}    
\int e^{-\beta^\mu p_\mu}\tilde{B}'(p^2)\tilde{\tau}'(p^2,b',q',s',|s'|)dp^4\;,    
\]    
\begin{equation}     
\hspace{1cm}=\frac{2\pi}{\beta}\int_0^{\infty} m\tilde{B}'(m^2) \tilde{\tau}'    
(m^2,\{\lambda\})K_1(\beta m)dm^2     
\end{equation}     
and $\varphi(\beta,\{\lambda\})$ the Laplace transform of the input    
term    
\begin{equation}    
\varphi(\beta,\{\lambda\})=    
\sum_{b'=-\infty}^{\infty}\lambda_b^{b'}    
\sum_{q'=-\infty}^{\infty}\lambda_q^{q'}    
\sum_{s'=-\infty}^{\infty}\lambda_S^{s'}    
\sum_{|s'|=0}^{\infty}\gamma_s^{|s'|}    
\int e^{-\beta^\mu p_\mu}g_{b'q's'|s'|}\tilde{B}'(p^2)\delta_0(p^2-   
m_{b'q's'|s'|}^2)    
dp^4\;.     
\end{equation}     
The masses $m_{bqs|s|}$ correspond to the masses of the input    
particles, which in this paper will be all the known    
hadrons with masses up to 2400 MeVs, the $g_{bqs|s|}$ are degeneracy factors    
due to spin and the $\lambda$'s are the fugacities of the input particles.    
Here we have used the extended version of SB [17], where the states are    
characterised by the set of fugacities relevant to baryon number $\lambda_b$,    
electric charge $\lambda_q$, strangeness $\lambda_s$ and partial    
strangeness equilibrium $\gamma_s$ (in the following we shall refer to this    
set of fugacities as $\{\lambda\}$, for short). The last fugacity $\gamma_s$    
is related to the number of strange quarks plus strange antiquarks [18], whereas    
the fugacity $\lambda_s$ is related to the number of strange quarks minus    
strange antiquarks. The introduction of $\gamma_s$ is necessary for the    
accurate theoretical prediction of the experimentally measured hadronic    
multiplicities.    
    
The bootstrap equation defines the boundaries of the hadronic phase    
since it exhibits a singularity at the point    
\begin{equation}    
\varphi(\beta,\{\lambda\}) = \ln4-1\;.    
\end{equation}     
    
From the physical point of view this singularity is connected with the behaviour    
of the mass spectrum as the mass tends to infinity    
\begin{equation}     
\tilde{\tau}'(m^2,\{\lambda\})\stackrel{m\rightarrow\infty}     
{\longrightarrow}     
2C(\{\lambda\})m^{-\alpha-1} \exp [m\beta^*(\{\lambda\})]\;,     
\end{equation}     
where $\beta=T^{-1}$ and $\beta^*$ corresponds to the inverse maximum    
temperature. After a certain point, as temperature rises, it is more preferable    
for the system to use the given energy in producing more hadronic states    
(since their number rises exponentially) than in increasing the kinetic energy    
of the already existing states. For our specific choice of  $\tilde{B}$ in (3) 
the   
exponent $\alpha=4$.   
    
In order to turn to the thermodynamics it is necessary to consider the    
fireball states in an external volume $V^{ext}$. This volume must     
be distinguished from the physical volume which is carried by each    
fireball.    
The partition function of the pointlike interacting hadrons is then    
\[    
\ln Z_{pt}(V^{ext},\beta,\{\lambda\})= \hspace{12cm}    
\]    
\[    
\sum_{b'=-\infty}^{\infty}\lambda_b^{b'}    
\sum_{q'=-\infty}^{\infty}\lambda_q^{q'}    
\sum_{s'=-\infty}^{\infty}\lambda_S^{s'}    
\sum_{|s'|=0}^{\infty}\gamma_s^{|s'|}    
\int \frac{2V^{ext}_{\mu}p^{\mu}}{{(2\pi)}^3}    
\tilde{\tau}'(p^2,b',q',s',|s'|) e^{-\beta^\mu p_\mu} dp^4\equiv    
\]    
\begin{equation}    
\int \frac{2V^{ext}_{\mu}p^{\mu}}{{(2\pi)}^3}    
\tilde{\tau}'(p^2,\{\lambda\}) e^{-\beta^\mu p_\mu} dp^4\;.    
\end{equation}    
Every choice for the mass spectrum in (3) (which leads to a certain exponent    
$\alpha$ in (8)) leads to a different partition function and so to a different    
physical behaviour of the system.    
The usual SB choice was $\alpha=2$, but more advantageous is our choice    
$\alpha=4$. With this choice a better physical behaviour is achieved as the     
system approaches the hadronic boundaries. Quantities like     
pressure, baryon density and energy density, even for     
point-like particles, no longer tend to infinity, as the system     
tends to the bootstrap singularity.    
It also allows for the bootstrap singularity to be reached in the     
thermodynamic limit [19], a necessity imposed by the Lee-Yang theory.    
Another point in favour of the choice $\alpha=4$ comes from the extension of    
SB to include strangeness [14,15].     
The strange chemical potential equals zero in the quark-gluon phase. With    
this particular choice of $\alpha$, $\mu_s$ acquires smaller positive values    
as the hadronic boundaries are approached.    
With the choice $\alpha=4$ the partition function can be written down and    
for point-like particles it assumes the form    
\begin{equation}    
\ln Z_{pt}(V^{ext},\beta,\{\lambda\})=    
\frac{4BV}{\beta^3}\int_{\beta}^{\infty} x^3 G(x,\{\lambda\})dx \equiv    
Vf(\beta,\{\lambda\})\;.    
\end{equation}     
For $\alpha=4$ the input term acquires the form    
\begin{equation}     
\varphi(\beta,\{\lambda\})=\frac{1}{2\pi^2 \beta B}     
\sum_{\rm a} (\lambda_{\rm a}(\{\lambda\})+\lambda_{\rm a}(\{\lambda\})^{-1})    
\sum_i g_{{\rm a}i} m_{{\rm a}i}^3 K_1 (\beta m_{{\rm a}i})\;\;,     
\end{equation}     
where the index ``${\rm a}$'' runs to all groups of hadrons, each of which    
is characterised by the same set of fugacities (e.g. Kaons with electric charge    
$Q=1$, $N$ and $\Delta$ Baryons with $Q=0$, etc.), ``$i$'' runs to all    
hadrons in the same group with different masses and $B$ is the energy    
density of the vacuum (MIT bag constant).    
    
By including corrections due to the finite size of hadrons (Van der    
Waals volume corrections) the repulsive part of the interaction is taken into    
account.    
The partition function for the real hadron gas is written [19,20] as    
follows    
\begin{equation}    
Z(V^{ext},\beta,\{\lambda\})=    
\sum_{N=0}^{\infty} \frac{1}{N\\!}    
\int \prod_{i=1}^{N}    
\left\{\frac{2(V^{ext}-p/4B)_{\mu}p_i^{\mu}}{{(2\pi)}^3} \right\}_{+}    
\tilde{\tau}'(p_i^2,\{\lambda\}) e^{-\beta_\mu p_i^\mu} dp_i^4\;.    
\end{equation}    
In the above relation the subscript $+$ indicates that each single bracket    
has to be non-negative, avoiding, thus, the negative contributions to the    
volume. The four momentum is $p^\mu=\sum_i^N p_i^\mu$ and    
because of its presence the integrations no longer factorize. The    
factorization property can be recovered through the grand canonical pressure    
partition function [19,20]    
\begin{equation}     
\pi(\beta,\xi,\{\lambda\})=    
\int_0^\infty dV e^{-\xi V} Z(V,\beta,\{\lambda\})\;,    
\end{equation}    
which is the Laplace transformed partition function with respect to volume.    
Provided that the thermodynamic limit    
$\lim_{V\rightarrow\infty}(1/V)\ln Z(\beta,V,\{\lambda\})$    
exists, the integral in (13) converges for the values    
\begin{equation}     
\xi > \xi_0(\beta,\{\lambda\}) \equiv \lim_{V\rightarrow\infty}    
\left[\frac{1}{V}\ln Z(\beta,V,\{\lambda\}) \right]\;.    
\end{equation}    
If we are constrained to values $\xi>\xi_0$ there is no need to employ    
Gaussian regularization [20] to evaluate (13) for $\xi\leq\xi_0$. Then    
the pressure partition function acquires the form    
\begin{equation}     
\pi(\xi,\beta,\{\lambda\})=\frac{1}{\xi-f(\beta+\xi/(4B),\{\lambda\})}\;,    
\end{equation}    
where $f=\ln Z_{pt}/V$.    
From (15) it is evident that the value of $\xi$ at the thermodynamic limit    
(14) can be obtained from    
\begin{equation}     
\xi_0=f(\beta+\xi_0/(4B),\{\lambda\}).    
\end{equation}    
We have, also, to note that for $\xi \neq 0$ the critical temperature of the    
hadronic state at zero density, $T_{0\;HG}$, does not depend only on the MIT    
bag constant, $B$, as is the case in [14-17], but on $\xi$, as well.    
    
The density and the pressure $P$ of the thermodynamic system can    
be obtained through the pressure grand canonical partition function (15)    
\begin{equation}    
\nu_{HG}(\xi,\beta,\{\lambda\})=\lambda    
\frac{\partial f(\beta+\xi/(4B),\{\lambda\})}{\partial \lambda}    
\left[1-\frac{1}{4B}    
\frac{\partial f(\beta+\xi/(4B),\{\lambda\})}{\partial \beta}\right]^{-1}\;,    
\end{equation}     
where $\lambda$ is the fugacity corresponding to the particular density, and    
\begin{equation}    
P_{HG}(\xi,\beta,\{\lambda\})=\frac{1}{\beta}    
f(\beta+\xi/(4B),\{\lambda\})    
\left[1-\frac{1}{4B}\frac{\partial f(\beta+\xi/(4B),\{\lambda\})}{\partial    
\beta}    
\right]^{-1}\;.    
\end{equation}     
    
Though volume is no longer an active variable of the system it can be    
calculated for given baryon density and $\nu_B$ (evaluated through (17)) and    
baryon number $<B>$ which is a conserved quantity. The volume    
would be retrieved through the relation    
\begin{equation}    
<V>=\frac{<B>}{\nu_B}\;.    
\end{equation}     
    
With the use of SB in order to describe interacting hadronic systems we    
can trace the possibility of a phase transition.     
The study of the pressure-volume isotherm curve is then necessary.     
When this curve is calculated a region of instability is revealed.    
In fact, this curve has a part (near the boundaries of the hadronic domain)    
where    
pressure decreases while volume decreases also (see Fig.~1).    
Such a behaviour is a signal of a {\it first order phase transition} which in    
turn can be mended with the use of a {\it Maxwell construction}.    
    
This behaviour is due to the formation of bigger and bigger clusters as the    
system tends to its boundaries in the phase diagram. In that way the effective    
number of particles    
is reduced, resulting, thus, to a decrease of pressure.    
This is the basic mechanism that will produce a first order transition at    
lower temperatures and a critical point at finite density.    
To show that this instability in the $P-V$ curve is the result of the    
attractive part of the interaction included in the SB we shall calculate a    
similar curve using the Ideal Hadron Gas (IHG) model with Van der Waals volume    
corrections (repulsive part of interaction). The logarithm of the partition    
function of IHG (corresponding to (10)) is    
\begin{eqnarray}     
\ln Z_{pt\;IHG}(V,\beta,\{\lambda\})&&\hspace{-0.1cm} \equiv    
Vf_{pt\;IHG}(\beta,\{\lambda\})= \nonumber \\    
\frac{V}{2\pi^2\beta}    
&&\hspace{-0.1cm}\sum_{\rm a} [\lambda_{\rm a}(\{\lambda\})+\lambda_{\rm    
a}(\{\lambda\})^{-    
1}]    
\sum_i g_{{\rm a}i} m_{{\rm a}i} K_2 (\beta m_{{\rm a}i})\;,     
\end{eqnarray}    
where $g_{{\rm a}i}$ are degeneracy factors due to spin and    
the index ${\rm a}$ runs to all groups of hadrons described by the same set    
of fugacities.    
This function can be used in eq.~(15) to calculate the Ideal Hadron Gas (IHG)    
pressure partition function in order to include Van der Waals volume    
corrections. The result is that the pressure is always found to increase as    
volume decreases, for constant temperature, exhibiting no region of    
instability and so no possibility of a phase transition.    
    
The comparison of SB with the IHG (with volume corrections) is displayed in    
Fig.~1, where $\nu_0$ is the normal nuclear density $\nu_0=0.14\;fm^{-3}$.    
In both cases (SB or IHG) the constraints $<S>=0$ (zero strangeness)    
and $<B>=2<Q>$ (isospin symmetric system, i.e. the net number of $u$ and $d$    
quarks are equal) have been imposed. Also strangeness is fully equilibrated    
which accounts to setting $\gamma_s=1$.

\vspace{2cm}        
    
{\large \bf 3. The partition function of Quark Matter}    
    
Having a description of the hadronic phase at hand, it is necessary to    
proceed with the thermodynamic behaviour of the quark-gluon phase.     
The QCD equation of state at finite temperature and baryon density calls for    
non-perturbative methods of approach. Lattice calculations have been    
performed but the power of such methods is limited by the sign and the    
overlap problems [21]. The overlap problem is treated with the    
reweighting method, called the Glasgow algorithm [22]. In [23] the    
method of imaginary chemical potential is used to solve the sign problem.    
In [24-26] various new methods have been used to tackle with the sign    
and/or the overlap problems. Especially in [26] the overlap problem is    
eliminated completely. In [27] lattice Taylor expansion is used around    
$\mu_B=0$, allowing the exploration of the phase transition of QCD at the    
low density regime.    
    
Lattice calculations of the pressure of the quark-gluon    
state have been performed at finite chemical potential in    
[4,6], using an improved reweighting technique. These publications include    
calculations for rather heavy $u$, $d$ quark masses.    
The mass of the $u$, $d$ quarks is about 65 MeV and the strange $s$    
quark 135 MeV [4,6]. The calculated pressure of the quark-gluon phase ($P/T^4$)    
at $\mu_B=0$    
is plotted against the ratio of temperature to the transition temperature    
of quark matter at zero baryon chemical potential $T/T_c$ in Fig.~2 of [4]. The    
temperature $T_c$ will be denoted as $T_{0\;QGP}$ in the following.    
The results of this graph are extrapolated to the continuum limit by    
multiplying the raw lattice results with a factor $c_p=0.518$ [4].    
    
The lattice calculations for finite chemical potential are    
summarised in Fig.~3 of [4], where the difference of pressure at non-zero    
chemical potential and the pressure at zero chemical potential    
($\Delta p/T^4 = [P(\mu \neq 0,T)-P(\mu = 0,T)]/T^4$) is plotted against    
$T/T_c$. Again the results of this graph are extrapolated to the continuum    
limit by multiplying the raw lattice results with a factor $c_{\mu}=0.446$ [4].    
In this graph five curves are plotted which correspond to baryon    
chemical potential of 100, 210, 330, 410, 530 MeV.    
    
With the use of Figs.~2, 3 in [4], it is possible to calculate in principle    
the pressure of the quark-gluon phase at any temperature and baryon chemical    
potential. The pressure is important, because knowledge of the    
pressure is equivalent to the knowledge of the partition     
function of the system in the grand canonical ensemble    
\begin{equation}    
\ln Z_{QGP}(V,T,\mu_B)=\frac{V}{T} P(T,\mu_B)    
\end{equation}    
    
In order to have a complete description of the dependence of the pressure on    
the temperature and the chemical potential we use two sets of fitting    
functions. For constant chemical potential the pressure as a function of    
$T/T_c$ is fitted through    
\begin{equation}    
f(x)=\frac{a_1}{x^{c_1} \left[ \exp\left(\frac{b_1}{x^{d_1}}\right)-   
1\right]^{f_1}}+    
\frac{a_2}{x^{c_2} \left[ \exp\left(\frac{b_2}{x^{d_2}}\right)-1\right]^{f_2}},    
\end{equation}    
where $a_i,b_i,c_i,d_i,f_i\;(i=1,2)$ depend on $\mu_B$, while for constant    
temperature the corresponding fit of the pressure as a function of $\mu_B$ is    
given by    
\begin{equation}    
g(x)=a+b \exp(c x^d),    
\end{equation}    
where $a,b,c,d$ depend on the temperature ratio $T/T_c$. The fitting    
procedure has to be performed in a self-consistent way and subsequently it is    
straightforward to    
evaluate the partition function, as well as its derivatives with respect to    
$\mu_B$ and $T$ at any given point $(T_1,\mu_{B\;1})$. In particular, to    
evaluate physical observables connected with the partition    
function and drive numerical routines the partial derivatives of the    
pressure up to second order with respect to temperature and fugacity have to    
be evaluated. These derivatives are then given in part A of the Appendix.    
    
In Fig.~2 we have reproduced the quark-gluon pressure as a function of the    
temperature for constant baryon chemical potential. The squares are points    
directly measured from the graphs of Fodor {\it et.~al} and the lines    
represent the calculation with the fits which has been performed on these    
points, via eq.~(22).    
Fig.~3 is a graph similar with Fig.~2, but we have focused on the area which    
is useful for our calculations, the area where the matching with the hadronic    
phase will be performed.    
Fig.~4 is a reproduction of the Fodor {\it et.~al} quark-gluon pressure as    
a function of the baryon chemical potential for constant temperature. The    
necessary fits have been performed with the use of eq.~(23).

\vspace{2cm}  
    
{\large \bf 4. The critical point in the phase diagram}    
    
After developing the necessary tools to handle the thermodynamic description    
for the quark-gluon phase as it is produced from the lattice, we can search    
for the possibility of a quark-hadron phase transition.    
    
First we have to deal with the free parameters that exist in our models.    
In [4,6] where the results of the pressure is presented the transition    
temperature of the quark state at zero density $T_0$ (called from now on    
$T_{0\;QGP}$) is not fixed.    
In [28], however, the QCD critical point is studied with quark mass input    
values closer to the physical ones and a zero-density temperature    
$T_{0\;QGP}=164 \pm 3$ MeV. Therefore in what follows, in order to limit the    
free parameters existing in our scheme, we choose    
\begin{equation}    
T_{0\;QGP}=164\;MeV\;.    
\end{equation}    
    
As far as the hadronic phase is concerned, an upper bound for the parameter     
$T_{0\;HG}$ can be fixed at the value $183$ MeV. This temperature allows for    
the best matching of the strange chemical potential $\mu_s$ between the   
hadronic and the QGP phase [15]. So we shall set     
\begin{equation}    
T_{0\;HG} \leq 183\;{\rm MeV}.    
\end{equation}    
    
The fact that $T_{0\;HG}$ and $T_{0\;QGP}$ acquire different values    
does not imply a contradiction.     
At $\mu_B=0$ the strongly interacting system belongs to the crossover
regime where the quark and hadron phases are indistinguishable.
Therefore, the zero-density parameters ($T_{0\;QGP},T_{0\;HG}$) do not
correspond physically to a distinct transition between the two phases;
they simply define the zero-density intercept of the extrapolated
critical line to small values of the chemical potential, beyond the
critical endpoint of the system at nonzero density. As a result, in the
region ($\mu_B=0$, 164 MeV$\leq T \leq 183$ MeV) of the phase diagram,
defined by the boundary values (24) and (25) a sharp thermodynamic
separation between the bootstrap and QCD phase may not exist. Although
the physics of the system in the crossover regime is not yet fully
understood, it is a plausible assumption (implied by (24), (25)) that
the two phases are distinguishable only in the domains: $\mu_B=0$,
$T\ll 164$ MeV (hadrons) and $T\gg 183$ MeV (quarks, gluons).

Turning now our attention to $\xi$, we adopt (14) in order to have always a    
real pressure partition function. For simplicity we also choose to have    
$\xi=const.$ for every set of $(T,\{\lambda\})$. This means that (14) has to    
be valid for every set of thermodynamic variables. Since the value of $\xi$ at    
the thermodynamic limit, $\xi_0$, depends on the choice of these variables    
we have to locate the specific set that gives us the highest value of    
$\xi_0$. For this reason we calculate $\xi_0$ on isotherms for different    
values of $\mu_B$ (fixing the remaining chemical potentials in order to fulfil    
the constraints $<B>=2<Q>$ and $<S>=0$). It is found that for constant    
temperature $\xi_0$ rises as a function of $\mu_B$. Then we calculate $\xi_0$    
on the maximum value of $\mu_B$ allowed for each temperature, i.e. on the    
critical curve (fixing accordingly the rest of the chemical potentials). It    
is found that on the bootstrap critical line $\xi_0$ rises as the temperature    
is increased. Thus the greatest value of $\xi_0$ corresponds to $T=T_{0\;HG}$    
and consequently $\{\lambda\}=\{1\}$. So in order to have a real pressure    
partition function for a constant value of $\xi$ all over the space of our    
thermodynamic variables it suffices to require    
\begin{equation}    
\xi > \xi_0(T_{0\;HG},\{\lambda\}=\{1\})\;.    
\end{equation}    
    
Finally as a consistency requirement on the thermodynamics of lattice QCD    
[28,29] and bootstrap matter we impose the constraint    
\begin{equation}    
T_{cr.p.} < T_{0\;QGP}\;.    
\end{equation}    
    
Then, if the values for the free parameters are chosen, within the above    
constraints, one may calculate    
for a specific temperature the pressure isotherms of Hadron Gas and QGP.    
Assuming that the baryon number is a conserved quantity to both phases, the    
equality of volumes is equivalent to the equality of baryon densities    
and so the connection of the isotherms of the two phases is possible through     
the relation    
\begin{equation}    
<V_{HG}>=<V_{QGP}>\Leftrightarrow    
\frac{<B_{HG}>}{\nu_{B\;HG}}=\frac{<B_{QGP}>}{\nu_{B\;QGP}}\Leftrightarrow    
\nu_{B\;HG}=\nu_{B\;QGP}    
\end{equation}    
The graph of the pressure-volume isotherm can be drawn using the plot of    
pressure against the inverse baryon density (see eq.~(19)).    
Then at the point where the isotherms of the two phases meet we have    
equal volumes for equal pressures.    
    
Tracing the point where the isotherms of two phases meet, we find     
that at a low temperature the intersection of QGP and SB pressure-volume    
isotherms takes place at a location where the Hadron Gas pressure is    
decreasing while volume    
decreases. The resulting pressure-volume curve includes an unstable part    
which has to be repaired through a suitable Maxwell construction. This    
curve includes a region where a first-order transition takes place.    
    
As the temperature increases, there exists a value for which the QGP    
and SB isotherms meet at a point where the Hadron Gas pressure has a maximum.    
In that case no Maxwell construction is     
needed and since this point is located at finite volume or not zero baryon    
density (equivalently not zero chemical potential) it can be associated with    
the QCD critical point.    
As temperature rises more, the resulting pressure-volume isotherm also    
increases while volume decreases without the need of a Maxwell construction    
and the situation belongs to the crossover region.    
    
A graph that summarises the situations met in the pressure volume isotherms    
of hadronic and quark systems in the neighbourhood of the critical point is    
Fig.~5. In this figure the hadronic isotherms    
have been calculated for parameters that fulfil the constraints (25)-(27)    
($T_{0\;HG} = 172$ MeV, MIT bag constant $B^{1/4} \simeq 136$ MeV,    
$\xi=177.1\cdot10^4\;{\rm MeV}^3$) while the quark-gluon isotherms are related    
to $T_{0\;QGP} = 164$ MeV. For the lower temperature isotherm a Maxwell    
construction is    
needed to remove the instability from the resulting curve. The horizontal    
line defines a partition with two equal surfaces (shaded) and represents the    
final pressure-volume curve after the completion of the Maxwell construction.     
At the temperature $T=162.06$ MeV the two curves meet at the point of    
maximum hadronic pressure and so a critical point is formed at finite volume.    
At a greater temperature the plotted pressure-volume isotherm corresponds to    
the crossover.    
    
To locate the critical point numerically with the use of the lattice    
partition function, for given parameters $\xi$, $T_{0\;HG}$ and $T_{0\;QGP}$,    
the conditions have to be determined for which the SB pressure is     
equal to the QGP pressure at the same volume,      
corresponding to the maximum SB pressure. Setting the factor of    
partial strangeness equilibrium $\gamma_s=1$ a hadronic state is    
characterised by the set of thermodynamic variables    
$(T, \lambda_u, \lambda_d, \lambda_s)$, while a quark-gluon state    
evaluated on the lattice [4] is characterised by the two variables    
$(T, \lambda_q')$. The $u$ and $d$ quarks are characterised by the same    
fugacity $\lambda_u'=\lambda_d'=\lambda_q'=\lambda_B'^{1/3}$. To    
evaluate the unknown variables we have to solve the following set of    
non-linear equations    
\begin{equation}    
\nu_{B\;SB}(T,\lambda_u,\lambda_d,\lambda_s)=    
\nu_{B\;QGP}(T,\lambda_q')    
\end{equation}    
\begin{equation}    
P_{SB}(T,\lambda_u,\lambda_d,\lambda_s)=    
P_{QGP}(T,\lambda_q')    
\end{equation}    
\begin{equation}    
\frac{dP_{SB}(T_1,\lambda_u,\lambda_d,\lambda_s)}{d V}=0    
\end{equation}    
\begin{equation}    
\left<\;S\;\right>_{SB}(T,\lambda_u,\lambda_d,\lambda_s)=0    
\end{equation}    
\begin{equation}    
\left<\;B\;\right>_{SB}(T,\lambda_u,\lambda_d,\lambda_s)-    
2\;\left<\;Q\;\right>_{SB}(T,\lambda_u,\lambda_d,\lambda_s)=0    
\end{equation}    
Eq.~(29) accounts for the equality of the baryon densities of the two phases    
which is equivalent to the equality of volumes, since the baryon number is a    
conserved quantity.    
Eq.~(30) is the equality of pressures of Hadron Gas and QGP.    
Eqs.~(29), (30) determine the point where the two pressure curves meet.    
Eq.~(31) requires that the meeting point of the two phases for a    
certain temperature is equal to the point which maximizes the Hadron Gas    
pressure and so this meeting point is the critical point. The form of this    
equation is discussed in detail in the Appendix.    
Eq.~(32) imposes strangeness neutrality in the hadronic phase.    
Eq.~(33) imposes isospin symmetry to the hadronic system, demanded in order    
to compare with the lattice studies.    
    
The area in the $(T,\mu_B)$ plane which gives solutions for the critical    
point    
compatible with the constraints (24)-(27) is depicted in Fig.~6. The line    
$\xi=\xi_{0\;max}$ represents solutions for the critical point with the    
requirement that $\xi$ is set to the thermodynamic limit value    
at $T=T_{0\;HG}$. Since the value of $\xi$ at thermodynamic limit is calculated    
through (16), we first determine $\xi$    
and the MIT bag constant $B$, for given $T_{0\;HG}$, by solving the    
following two equations    
\begin{equation}    
\xi=f(1/T_{0\;HG}+\xi/4B,\{\lambda\}=\{1\};B)\;.    
\end{equation}    
\begin{equation}    
\varphi(1/T_{0\;HG}+\xi/4B,\{\lambda\}=\{1\};B)=\ln4-1\;.    
\end{equation}    
The last equation determines the critical hadronic curve at zero density.    
We, then, insert the calculated values of $\xi$ and    
$B$, for the given value of $T_{0\;HG}$, in the system of eqs.~(29)-(33),    
which is solved to extract the thermodynamic variables that locate the    
critical point on the phase diagram. All the points on the right of the    
$\xi=\xi_{0\;max}$ curve on the $T-\mu_B$ plane are compatible with the    
requirement (26).    
    
The compatible area is also limited by the curve of constant $T_{0\;HG}$,    
according to (25). Setting $T_{0\;HG}$ and choosing a value of $\xi$ that    
fulfils (26) we fist solve (35) to determine $B$. Then we solve the system    
(29)-(33) to draw the line of constant $T_{0\;HG}$.    
    
In Fig.~6 we have also drawn the line $T=164$ MeV, which excludes solutions    
that do not fulfil (27). The shaded area represents all the points that    
qualify as a solution for the critical point, compatible with eqs.~(24)-(27).  
    
Recent lattice QCD studies offer, apart from the quark-gluon pressure which    
has been a basic ingredient in our approach, important results on the    
existence and location of the critical point itself.    
In [29] the critical point is found to reside at $T_{cr.p.}=160\pm3.5$ MeV and    
$\mu_B=725\pm35$ MeV, with $T_{0\;QGP}=172\pm3$ MeV. These    
calculations have the drawback that have been performed with $u$ and $d$    
quark mass which has a value about four times the physical value.    
Improved calculations have been performed in [28], where the light quark    
masses have decreased by a factor of 3 down their physical values. The    
critical point is found now (with $T_{0\;QGP}=164\pm3$ MeV) to be at    
$T_{cr.p.}=162\pm2$ MeV and $\mu_B=360\pm40$ MeV, a value which is    
considerably reduced with respect to the previous one. This point is depicted    
on Fig.~6 with the full star and it falls completely inside the compatible    
domain of the critical point, according to our calculations.    
In [27] the critical point has also been determined using Taylor    
expansion around $\mu_B=0$ of the 3-flavour QCD equation of state. The    
critical point is found to reside at $\mu_B=420$ MeV.    
In [30] a method is exhibited about how to locate the critical point    
using imaginary values of $\mu_B$. This study also shows the sensitivity of    
the critical point on the strange quark mass.    
    
In Figs.~7-8 we illustrate a representative solution for the critical point    
as well as for the QGP-hadron transition line from those which are    
included in the shaded area of Fig.~6 on the $(T,\mu_B)$ plane.    
The chosen critical point is associated with $T_{0\;HG}=172$ MeV, while the    
rest of the parameters are those of Fig.~5. It is located at    
$T_{cr.p.}=162.1$ MeV and $\mu_B=218.7$ MeV.    
The full circle represents our solution for the critical point. This circle    
is the endpoint of the solid thick line representing the bootstrap    
calculation of the maximum hadronic pressure which is close to the first    
order critical line.    
    
In Fig.~7 we also compare our solutions for the critical point with other    
calculations. The full stars represent solutions of lattice QCD. The    
points depicted as ``lattice reweighting I and II'' are the points from    
[29] and [28] respectively and the one depicted as ``Taylor expansion'' is taken    
from [27].    
The solid thin line is the first order transition line evaluated in [29] with    
large $u$, $d$ quark masses. The dotted line represents the crossover in this    
calculation.    
The two slashed lines represent the magnitude of the slope    
$d^2 T/d\mu^2$ taken from lattice Taylor expansion and which    
indicate the phase transition line according to [27].      
Theoretical predictions on the location of the critical point from various    
models [3,31-36] are depicted as open squares in Fig.~7. The labels    
adopted are those of Table I in [21]. The open circle in Fig.~7 is the    
prediction of the critical point within the statistical bootstrap    
($\xi=0$) from [8], where a simplified partition function (MIT bag) for the    
quark phase was used.    
    
In Figs.~6 and 8 we compare our solutions for the critical point with the    
freeze-out points from different experiments. In Fig.~6 we depict freeze-out    
points from NA49 at 158 AGeV [37] for systems of different size ($C+C$,    
$Si+Si$, $Pb+Pb$) that fall inside the compatible domain    
of the critical point, if the errors in their determination are taken into    
account. These experiments are interesting since they could trace critical    
fluctuations associated with a critical point of second order [44].    
In Fig.~8 the comparison is extended to a larger number of experiments    
[38-43]. On the same graph we have also depicted the curve of $<E>/<N>=1$ GeV    
[45] (reproduced from [46]) that fits freeze-out points which are spread to   
a wide region of the phase diagram.    
It is evident that our calculations set the critical point to a location    
easily accessible by experiments, especially by CERN/SPS.    
    
\newpage                                        
\vspace{1cm}    
    
{\large \bf 5. Conclusions}    
    
Statistical Bootstrap presents a more accurate description of the    
hadronic phase than the ideal Hadron Gas, since it includes in a self    
consistent way the attractive part of the interaction among hadrons through    
the mass spectrum, as well as the repulsive part through Van der Waals volume    
corrections. This interaction is crucial to    
investigate critical phenomena in connection with the state of quark-gluon    
plasma. Among the predictions of the bootstrap model is the limitation of the    
hadronic phase and the forming of an instability in the pressure-volume    
isotherm near the hadronic boundaries. This instability is connected    
to a first order quark-hadron phase transition and the existence of a    
critical endpoint in the strongly interacting matter.    
    
A more realistic pressure diagram of the quark-gluon phase is available from    
lattice calculations, despite the fact that unphysical values of the light    
quark masses are still involved in these studies. From these results the    
lattice partition function of the    
quark state, as well as, all the necessary derivatives can be calculated,    
allowing the evaluation of any physical observable.    
    
The joining of the SB and the lattice partition function for the haronic and    
the quark state respectively, allows for the determination of a critical    
point at finite baryon chemical potential which can be related to the    
critical point of QCD.    
    
More recent lattice calculations [28] drive the position of the critical point    
to smaller values of baryon chemical potential as the values of $u$, $d$ quark    
masses approach their physical values. It is interesting that the current    
location is situated in the $(T,\mu_B)$ plane in a region easily accessible    
by the freeze-out conditions of experiments at the CERN/SPS.    
    
Setting the free parameters in our model in a way to fulfil certain    
constraints we are left with a compatible domain in the $(T,\mu_B)$ plane    
for the location of the critical point. Recent lattice calculations    
[28] drive the critical point within the domain of our solutions.    
    
In a previous work [9,8] a similar solution was found with the use of a    
simplified partition function for the quark-gluon system, based on    
the MIT bag model. Therefore, the basic mechanism in our approach for the    
formation of a critical point in the strongly interacting matter is not    
associated with the details of the quark-gluon partition function but mainly    
with the instability of hadronic matter revealed by Statistical Bootstrap.    
In particular the local maximum of pressure in the $P-V$ diagram (Fig.~5) of    
hot hadronic matter lies in the origin of the formation mechanism of the    
critical endpoint.

\vspace{1cm}    
    
{\large \bf Acknowledgements}    
    
This work was supported in part by the Research Committee of the University    
of Athens and the EPEAEK research funding program ``Pythagoras I''
(70/3/7315).
One of us (N.~G.~A.) wishes to thank Larry McLerran and Dmitri Kharzeev for    
the hospitality extended to him at the Brookhaven National Laboratory.

\vspace{1cm}    
    
{\large \bf Appendix}    
    
{\large \bf A.} The fitting procedures on curves of constant temperature and    
chemical potential allow us to evaluate the derivatives with respect to $T$    
for constant $\mu_B$ and with respect to $\mu_B$ for constant $T$. Physical    
observables, however, are given as derivatives with respect to temperature    
for constant fugacity or with respect to $\lambda_B$ for constant $T$. The    
evaluation of the latter (for the pressure) is given by    
\[    
\hspace{3.8cm}    
\left. \frac{\partial P}{\partial \lambda_B}\right|_T=    
\left. \frac{\partial P}{\partial \mu_B}\right|_T \frac{T}{\lambda_B} ,    
\hspace{3cm} (A.1)    
\]    
\[    
\hspace{3.8cm}    
\left. \frac{\partial P}{\partial T}\right|_{\lambda_B}=    
\left. \frac{\partial P}{\partial T}\right|_{\mu_B}+    
\left. \frac{\partial P}{\partial \lambda_B}\right|_{T}    
\frac{\lambda_B \mu_B}{T^2}=    
\left. \frac{\partial P}{\partial T}\right|_{\mu_B}+    
\left. \frac{\partial P}{\partial \mu_B}\right|_{T}    
\frac{\mu_B}{T},    
\hspace{1cm} (A.2)    
\]    
\[    
\hspace{3.8cm}    
\left. \frac{\partial^2 P}{\partial \lambda_B^2}\right|_{T}=    
- \left. \frac{\partial P}{\partial \mu_B}\right|_{T} \frac{T}{\lambda_B^2}+    
\left. \frac{\partial^2 P}{\partial \mu_B^2}\right|_{T}    
\frac{T^2}{\lambda_B^2},    
\hspace{3cm} (A.3)    
\]    
\[    
\hspace{3.8cm}    
\left. \frac{\partial^2 P}{\partial T^2}\right|_{\lambda_B}=    
\left. \frac{\partial^2 P}{\partial T^2}\right|_{\mu_B} +    
\frac{\partial^2 P}{\partial \mu_B \partial T} \frac{2 \mu_B}{T} +    
\left. \frac{\partial^2 P}{\partial \mu_B^2}\right|_{T} \frac{\mu_B^2}{T^2},    
\hspace{3cm} (A.4)    
\]    
\[    
\hspace{3.8cm}    
\frac{\partial^2 P}{\partial \lambda_B \partial T}=    
\frac{\partial^2 P}{\partial T \partial \mu_B} \frac{T}{\lambda_B}+    
\left. \frac{\partial^2 P}{\partial \mu_B^2} \right|_{T}    
\frac{\mu_B}{\lambda_B} +    
\left. \frac{\partial P}{\partial \mu_B}\right|_{T} \frac{1}{\lambda_B}.    
\hspace{3cm} (A.5)    
\]    
In eqs.~(A.4), (A.5) where the 2nd partial derivative with respect to two    
different variables of $P$ appears, the pressure is considered as a function of    
these two variables.

{\large \bf B.} After choosing specific values for the parameters $\xi$ and    
$B$, the requirement of eq.~(31) has to be fulfilled for a certain temperature    
$T=T_1$ and in the presence of two constraints    
$g_1 \equiv \left<\;S\;\right>_{SB}(T_1,\lambda_u,\lambda_d,\lambda_s)=0$    
and $g_2 \equiv$    
$\left<\;B\;\right>_{SB}(T_1,\lambda_u,\lambda_d,\lambda_s)-    
2\;\left<\;Q\;\right>_{SB}(T_1,\lambda_u,\lambda_d,\lambda_s)=0$.    
Since the maximum of pressure is found for a certain isotherm, the    
temperature $T_1$ may not be considered in the following as an active    
variable. The    
same is true for $\xi$ and $B$, since the maximum pressure is evaluated for    
constant values of these parameters. With the above considerations and    
eq.~(19) we may write    
\[    
\frac{dP_{SB}(T_1,\lambda_u,\lambda_d,\lambda_s)}{d V}=0 \Rightarrow    
-\frac{<B>}{\nu_B^2}    
\frac{dP_{SB}(T_1,\lambda_u,\lambda_d,\lambda_s)}{d \nu_B}=0 \Rightarrow    
\]    
\[    
\frac{dP_{SB}(T_1,\lambda_u,\lambda_d,\lambda_s)}{d \nu_B}=0    
\Rightarrow dP_{SB}=0 \Rightarrow    
\frac{\partial P_{SB}}{\partial \lambda_u} d\lambda_u+    
\frac{\partial P_{SB}}{\partial \lambda_d} d\lambda_d+    
\frac{\partial P_{SB}}{\partial \lambda_s} d\lambda_s=0\Rightarrow    
\]    
\[    
\hspace{3.8cm}    
\frac{\partial P_{SB}}{\partial \lambda_u} +    
\frac{\partial P_{SB}}{\partial \lambda_d} \frac{d\lambda_d}{d\lambda_u}+    
\frac{\partial P_{SB}}{\partial \lambda_s} \frac{d\lambda_s}{d\lambda_u}=0    
\hspace{3cm} (B.1)    
\]    
    
As far the constraint $g_1$ is concerned, we have    
\[    
g_1(T_1,\lambda_u,\lambda_d,\lambda_s)=0\Rightarrow dg_1=0\Rightarrow    
\frac{\partial g_1}{\partial \lambda_u} d\lambda_u+    
\frac{\partial g_1}{\partial \lambda_d} d\lambda_d+    
\frac{\partial g_1}{\partial \lambda_s} d\lambda_s=0\Rightarrow    
\]    
\[    
\hspace{3.8cm}    
\frac{\partial g_1}{\partial \lambda_d} \frac{d\lambda_d}{d\lambda_u}+    
\frac{\partial g_1}{\partial \lambda_s} \frac{d\lambda_s}{d\lambda_u}=    
-\frac{\partial g_1}{\partial \lambda_u}    
\hspace{3cm} (B.2)    
\]    
Similarly for the constraint $g_2$ we have    
\[    
\hspace{3.8cm}    
\frac{\partial g_2}{\partial \lambda_d} \frac{d\lambda_d}{d\lambda_u}+    
\frac{\partial g_2}{\partial \lambda_s} \frac{d\lambda_s}{d\lambda_u}=    
-\frac{\partial g_2}{\partial \lambda_u}    
\hspace{3cm} (B.3)    
\]    
Eqs.~(B.2) and (B.3) may considered as a system of two equations which    
can be solved to determine $d\lambda_d / d\lambda_u$ and    
$d\lambda_s / d\lambda_u$    
\[    
\hspace{3.8cm}\frac{d\lambda_d}{d\lambda_u}=\frac{1}{D}\left(    
\frac{\partial g_1}{\partial \lambda_s}    
\frac{\partial g_2}{\partial \lambda_u}-    
\frac{\partial g_2}{\partial \lambda_s}    
\frac{\partial g_1}{\partial \lambda_u} \right), \hspace{3cm} (B.4)    
\]    
\[    
\hspace{3.8cm}\frac{d\lambda_s}{d\lambda_u}=\frac{1}{D}\left(    
\frac{\partial g_1}{\partial \lambda_u}    
\frac{\partial g_2}{\partial \lambda_d}-    
\frac{\partial g_2}{\partial \lambda_u}    
\frac{\partial g_1}{\partial \lambda_d} \right), \hspace{3cm} (B.5)    
\]    
with    
\[    
\hspace{3.8cm}D=\frac{\partial g_1}{\partial \lambda_d}    
\frac{\partial g_2}{\partial \lambda_s}-    
\frac{\partial g_2}{\partial \lambda_d}    
\frac{\partial g_1}{\partial \lambda_s}. \hspace{3cm} (B.6)    
\]    
    
Eqs.~(B.4) and (B.5) may now be inserted to eq.~(B.1) to give    
\[    
\frac{\partial P_{SB}}{\partial \lambda_u} +    
\frac{\partial P_{SB}}{\partial \lambda_d}    
\left(    
\frac{\partial g_1}{\partial \lambda_s}    
\frac{\partial g_2}{\partial \lambda_u}-    
\frac{\partial g_2}{\partial \lambda_s}    
\frac{\partial g_1}{\partial \lambda_u} \right) \frac{1}{D} +    
\frac{\partial P_{SB}}{\partial \lambda_s}    
\left(    
\frac{\partial g_1}{\partial \lambda_u}    
\frac{\partial g_2}{\partial \lambda_d}-    
\frac{\partial g_2}{\partial \lambda_u}    
\frac{\partial g_1}{\partial \lambda_d} \right) \frac{1}{D}    
=0 \hspace{0.3cm} (B.7)    
\]    
Eq.~(B.7) is the form of eq.~(31) for the maximum hadron pressure in a    
certain isotherm. The main contribution comes from the term    
$\partial P_{SB} / \partial \lambda_u$,    
as it is verified by comparing the maximum pressure found in the present    
work using the full equation (B.7) with the solution found using only the    
first term in [9,8].

\vspace{1cm}    
    
{\large \bf Figure Captions}    
    
\newtheorem{f}{Fig.}    
\begin{f}     
\rm Isotherm pressure-volume curve for SB and IHG (both with Van der Waals    
    volume corrections using the pressure ensemble with the same value of    
    $\xi$).  The SB curve is    
    drawn for $T_{0\;HG}=172$ MeV ($B^{1/4}\simeq136$ MeV).    
\end{f}     
\begin{f}    
\rm The pressure of the quark-gluon state divided by $T^4$ versus the ratio    
    $T/T_{0\;QGP}$, for constant baryon chemical potential. The lines from    
    bottom to top correspond to gradually increasing values of $\mu_B$. The    
    squares represent direct measurement from the Figs.~2, 3 in [4] which depict    
    the lattice calculation and the lines indicate our fits to these points.    
\end{f}    
\begin{f}    
\rm Similar graph with Fig.~2. The pressure of the quark-gluon state is    
    divided by $T_{0\;QGP}^4$. The plot focuses on the region where    
    the matching with the hadronic state will take place.    
\end{f}    
\begin{f}    
\rm The pressure of the quark-gluon state divided by $T_{0\;QGP}^4$ versus    
    the baryon chemical potential, for constant values of the ratio    
    $T/T_{0\;QGP}$. The squares represent direct measurement from Figs.~2, 3    
    in [4] which depict the lattice calculation. The lines indicate our fits    
    to these points.    
\end{f}    
\begin{f}    
\rm Three isotherm pressure-volume curves for Hadron Gas (using SB) and QGP    
    phase (using the lattice pressure of [4]).    
    The low temperature isotherm    
    needs Maxwell construction, the middle temperature isotherm develops a    
    critical point and the high temperature isotherm corresponds to    
    crossover.    
\end{f}     
\begin{f}     
\rm Domain in the $(T,\mu_B)$ plane (shaded area) which is compatible with    
    solutions for the position of the critical point, after imposing the    
    constraints (24)-(27) on the free parameters. There are, also, displayed    
    the freeze-out points from NA49 at 158 AGeV [37] and the lattice calculated    
    critical point from [28].    
\end{f}    
\begin{f}    
\rm The critical point domain of Fig.~6 and a representative solution of the    
    critical point (full circle-``SB/II'') at the $(T,\mu_B)$ plane, as well    
    as the first order part of    
    the quark-hadron transition (solid thick line). The stars represent the    
    lattice calculated critical points [27,28,29]. The line labelled    
    ``lat. rew.'' represents the phase boundaries from [29] and the lines    
    labelled ``Tayl. exp.'' represent the phase boundaries from [27]. The    
    open squares are theoretical calculations for the critical point [3,31-36].    
    The open circle (``SB/I'') is the calculation of critical point using    
    bootstrap model in [8].    
\end{f}    
\begin{f}    
\rm Comparison of the positions of our critical point solutions presented    
    in Figs.~6-7, with the freeze-out points from different experiments [38-43].    
    It is, also, displayed, with slashed line the $<E>/<N>=1$ GeV    
    freeze-out curve of [45], reproduced from [46].    
\end{f}

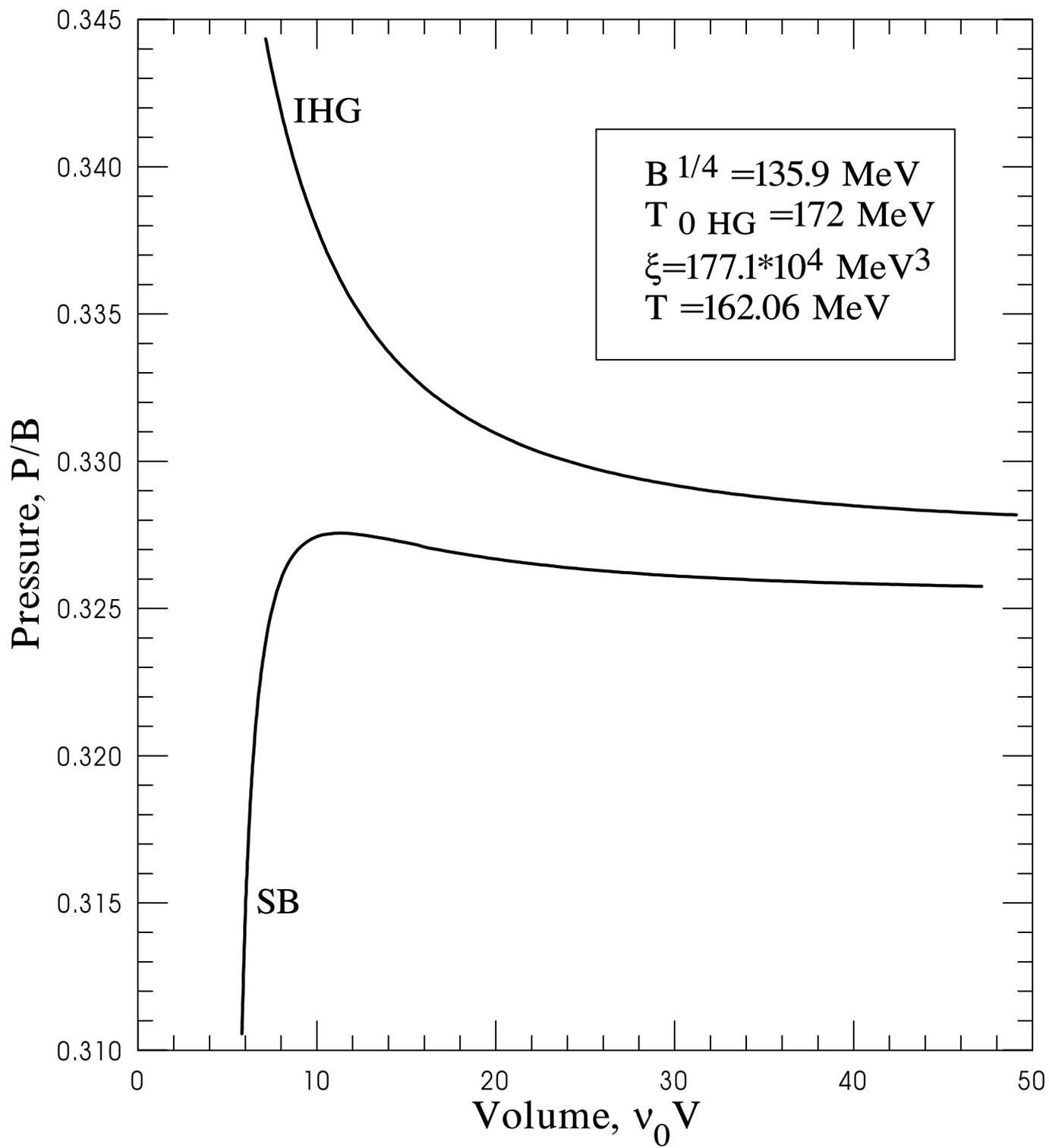
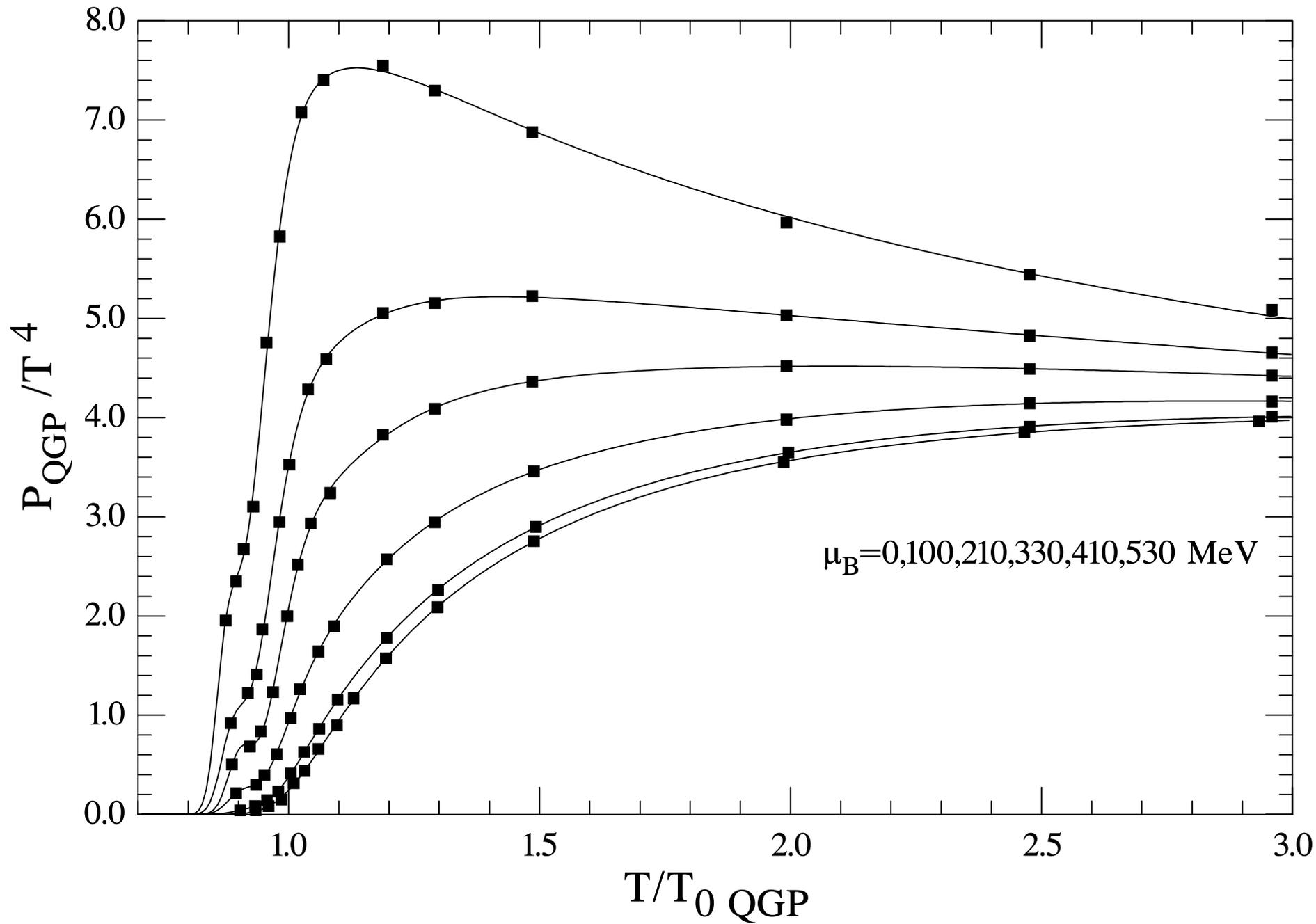
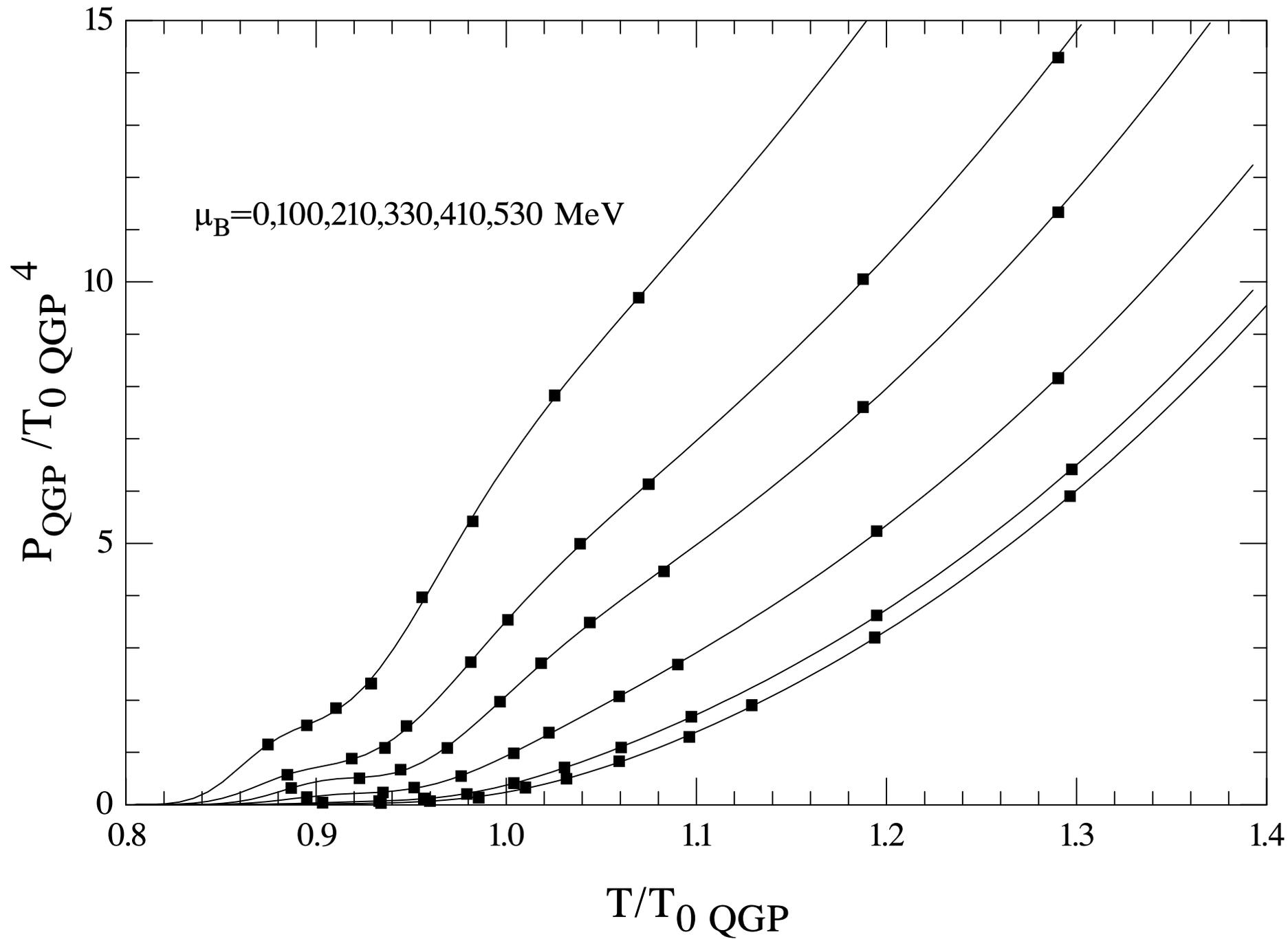
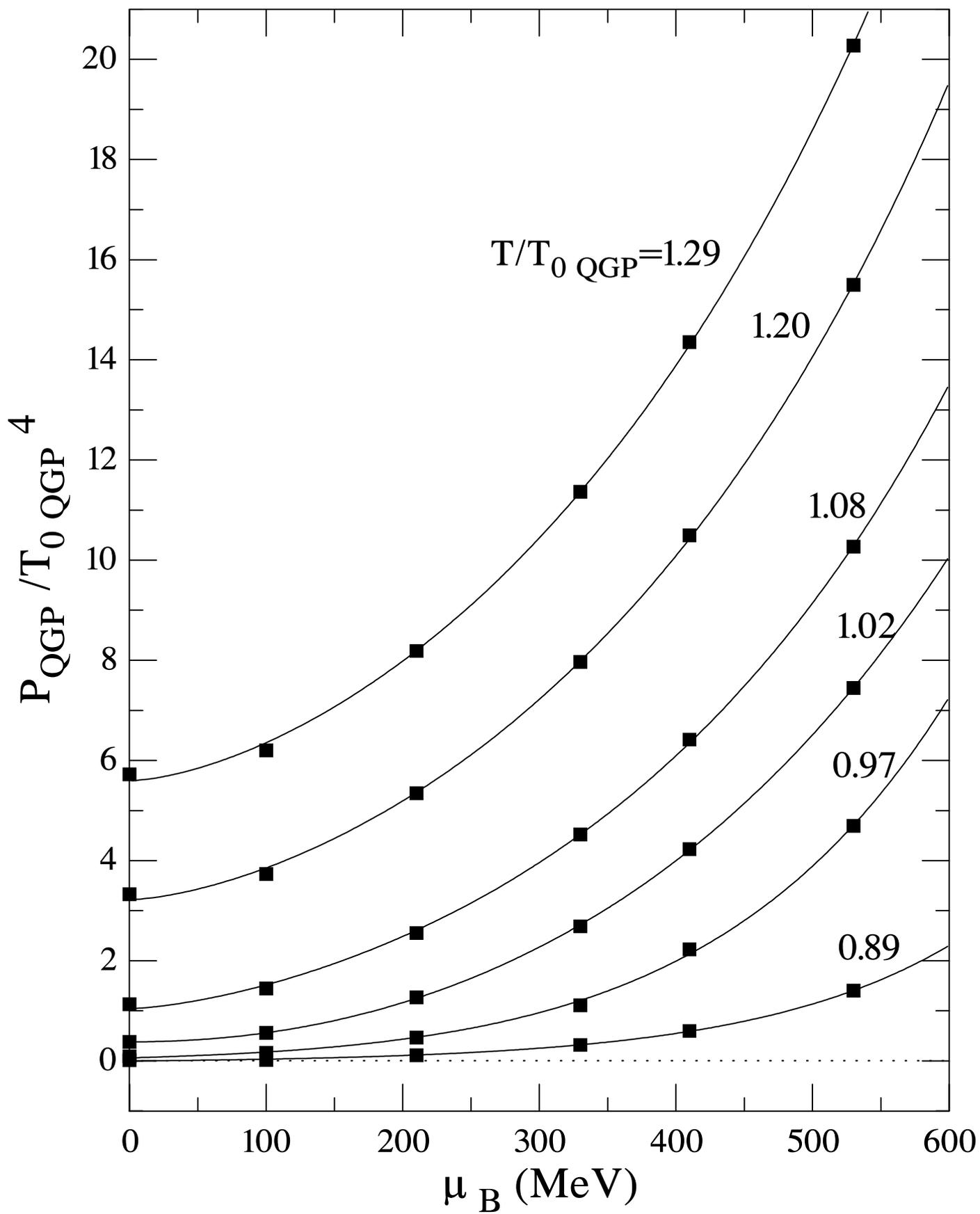
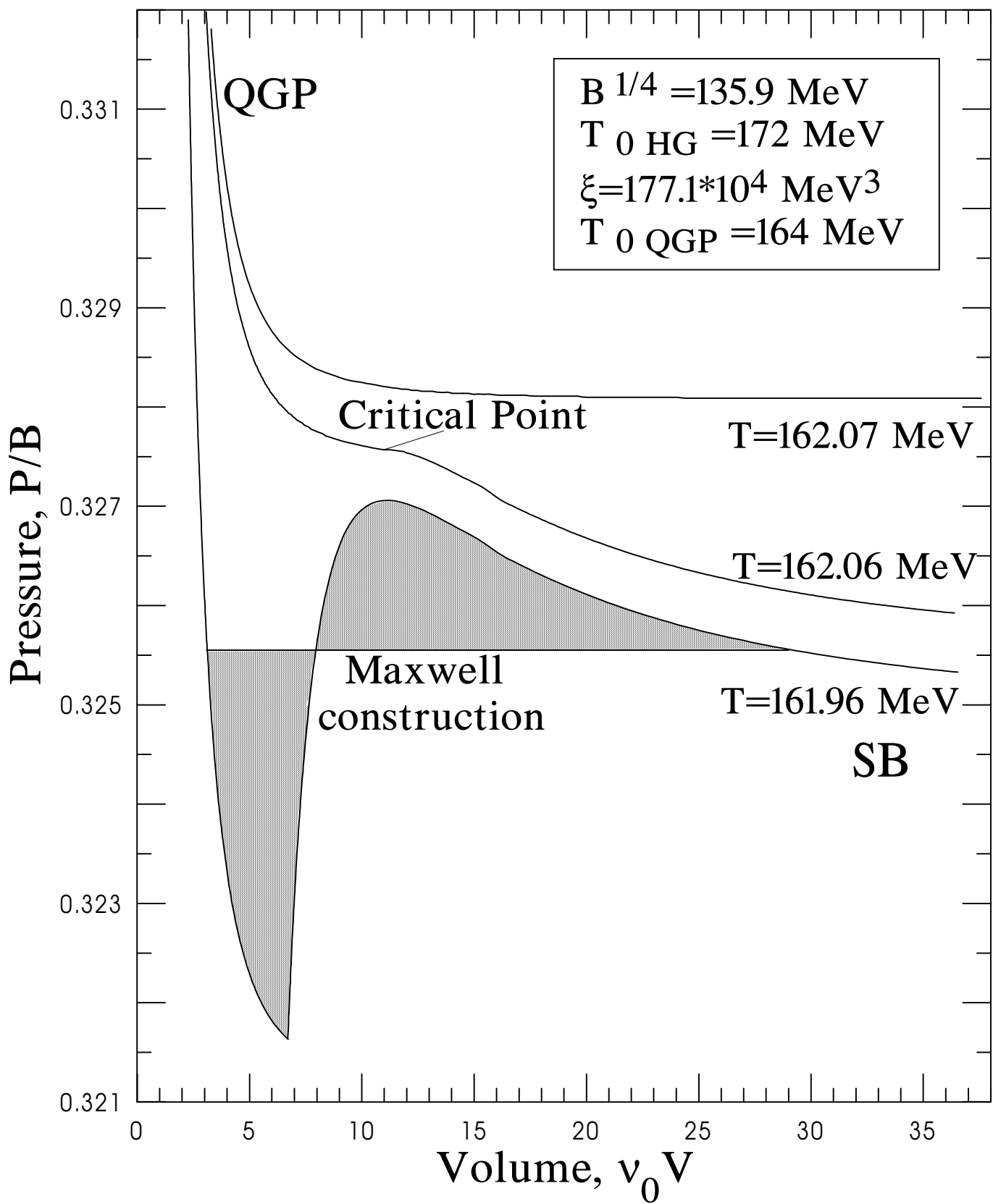
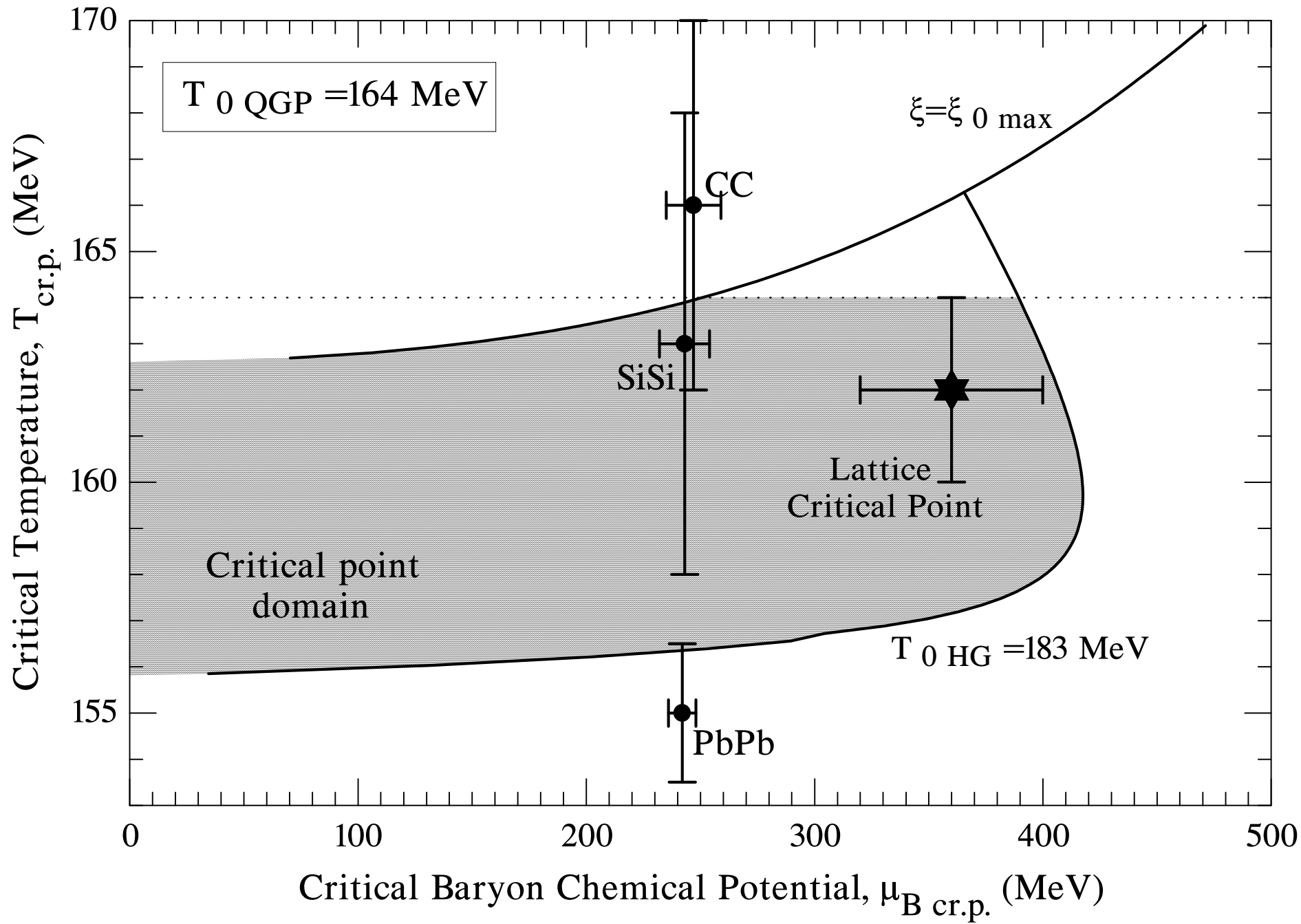
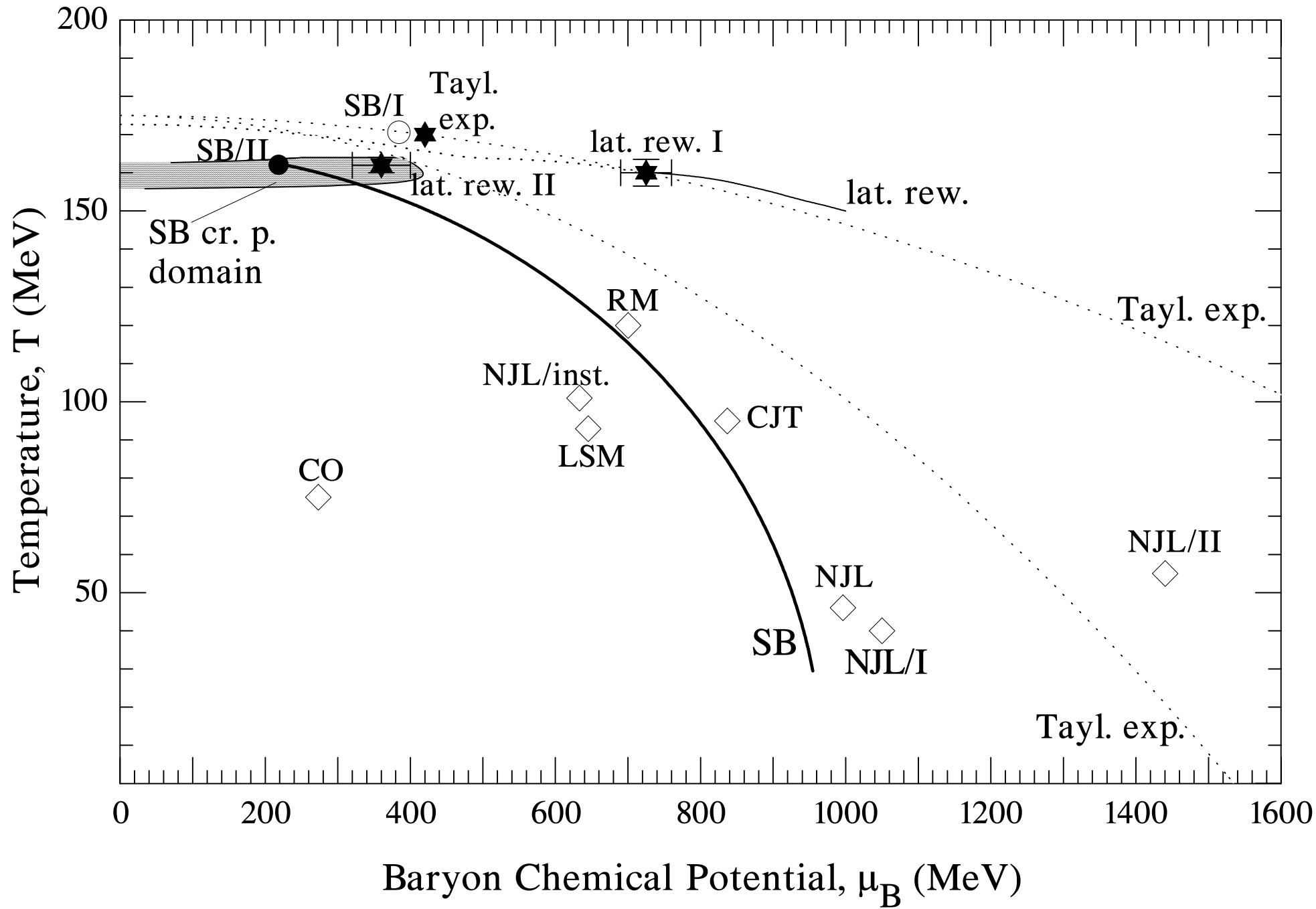
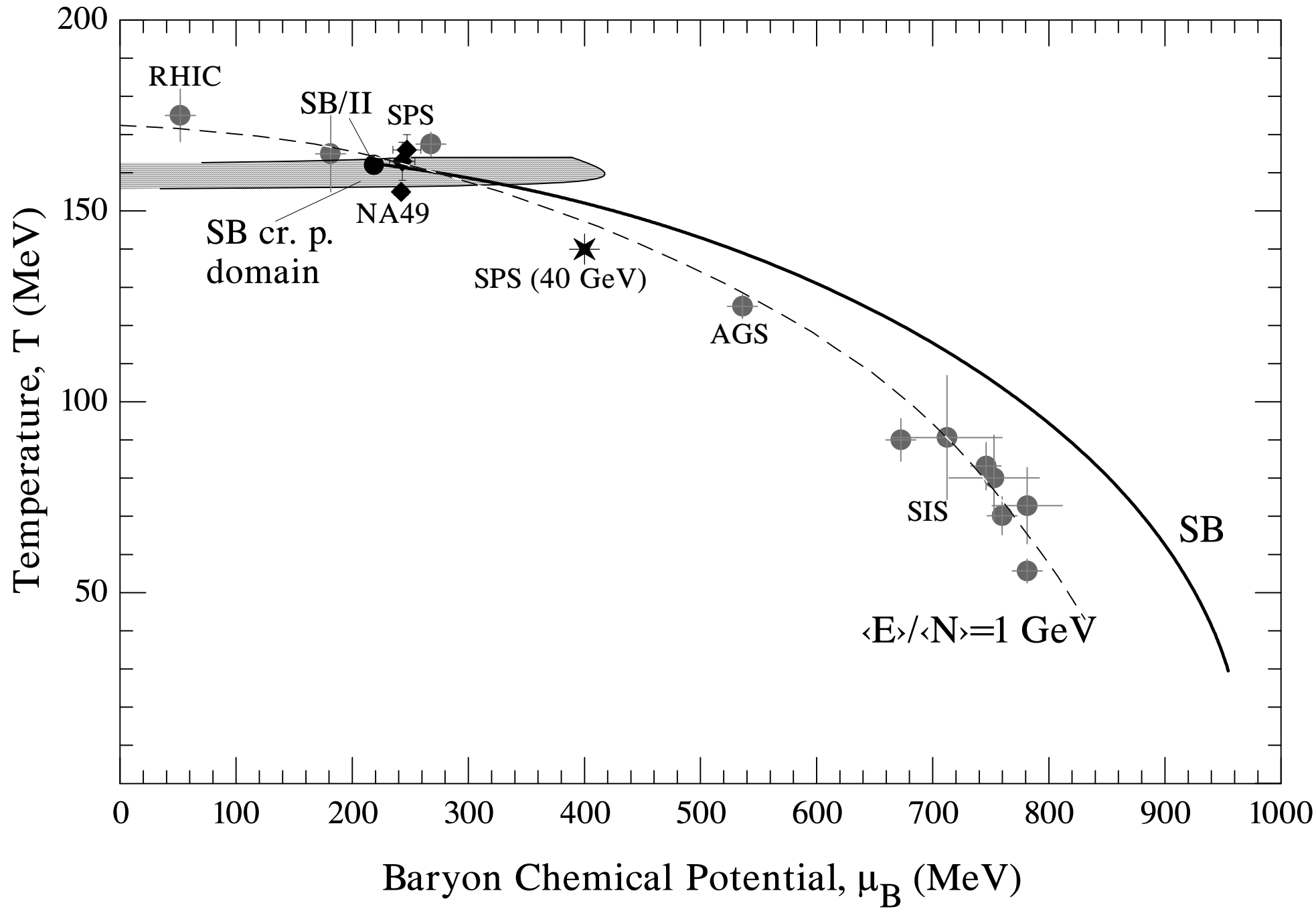
\end{document}